\documentclass[twoside,twocolumn,9pt]{article}
\usepackage{extsizes}
\usepackage[super,sort&compress,comma]{natbib} 
\usepackage[version=3]{mhchem}
\usepackage[left=1.5cm, right=1.5cm, top=1.785cm, bottom=2.0cm]{geometry}
\usepackage{balance}
\usepackage{mathptmx}
\usepackage{sectsty}
\usepackage{graphicx} 
\usepackage{lastpage}
\usepackage[format=plain,justification=justified,singlelinecheck=false,font={stretch=1.125,small,sf},labelfont=bf,labelsep=space]{caption}
\usepackage{float}
\usepackage{fancyhdr}
\usepackage{fnpos}
\usepackage[english]{babel}
\addto{\captionsenglish}{%
  \renewcommand{\refname}{Notes and references}
}
\usepackage{array}
\usepackage{droidsans}
\usepackage{charter}
\usepackage[T1]{fontenc}
\usepackage[usenames,dvipsnames]{xcolor}
\usepackage{setspace}
\usepackage[compact]{titlesec}
\usepackage{hyperref}
\usepackage[frozencache,cachedir=cache]{minted}

\usepackage{epstopdf}

\definecolor{cream}{RGB}{222,217,201}

\begin{document}

\pagestyle{fancy}
\thispagestyle{plain}
\fancypagestyle{plain}{
\renewcommand{\headrulewidth}{0pt}
}

\makeFNbottom
\makeatletter
\renewcommand\LARGE{\@setfontsize\LARGE{15pt}{17}}
\renewcommand\Large{\@setfontsize\Large{12pt}{14}}
\renewcommand\large{\@setfontsize\large{10pt}{12}}
\renewcommand\footnotesize{\@setfontsize\footnotesize{7pt}{10}}
\makeatother

\renewcommand{\thefootnote}{\fnsymbol{footnote}}
\renewcommand\footnoterule{\vspace*{1pt}%
\color{cream}\hrule width 3.5in height 0.4pt \color{black}\vspace*{5pt}} 
\setcounter{secnumdepth}{5}

\makeatletter 
\renewcommand\@biblabel[1]{#1}            
\renewcommand\@makefntext[1]%
{\noindent\makebox[0pt][r]{\@thefnmark\,}#1}
\makeatother 
\renewcommand{\figurename}{\small{Fig.}~}
\sectionfont{\sffamily\Large}
\subsectionfont{\normalsize}
\subsubsectionfont{\bf}
\setstretch{1.125} 
\setlength{\skip\footins}{0.8cm}
\setlength{\footnotesep}{0.25cm}
\setlength{\jot}{10pt}
\titlespacing*{\section}{0pt}{4pt}{4pt}
\titlespacing*{\subsection}{0pt}{15pt}{1pt}

\fancyfoot{}
\fancyfoot[LO,RE]{\vspace{-7.1pt}\includegraphics[height=9pt]{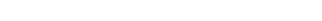}}
\fancyfoot[CO]{\vspace{-7.1pt}\hspace{13.2cm}\includegraphics{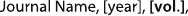}}
\fancyfoot[CE]{\vspace{-7.2pt}\hspace{-14.2cm}\includegraphics{head_foot/RF}}
\fancyfoot[RO]{\footnotesize{\sffamily{1--\pageref{LastPage} ~\textbar  \hspace{2pt}\thepage}}}
\fancyfoot[LE]{\footnotesize{\sffamily{\thepage~\textbar\hspace{3.45cm} 1--\pageref{LastPage}}}}
\fancyhead{}
\renewcommand{\headrulewidth}{0pt} 
\renewcommand{\footrulewidth}{0pt}
\setlength{\arrayrulewidth}{1pt}
\setlength{\columnsep}{6.5mm}
\setlength\bibsep{1pt}

\makeatletter 
\newlength{\figrulesep} 
\setlength{\figrulesep}{0.5\textfloatsep} 

\newcommand{\topfigrule}{\vspace*{-1pt}%
\noindent{\color{cream}\rule[-\figrulesep]{\columnwidth}{1.5pt}} }

\newcommand{\botfigrule}{\vspace*{-2pt}%
\noindent{\color{cream}\rule[\figrulesep]{\columnwidth}{1.5pt}} }

\newcommand{\dblfigrule}{\vspace*{-1pt}%
\noindent{\color{cream}\rule[-\figrulesep]{\textwidth}{1.5pt}} }

\makeatother

\twocolumn[
  \begin{@twocolumnfalse}
{\includegraphics[height=30pt]{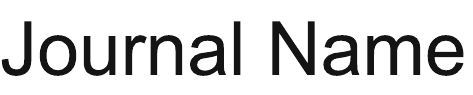}\hfill\raisebox{0pt}[0pt][0pt]{\includegraphics[height=55pt]{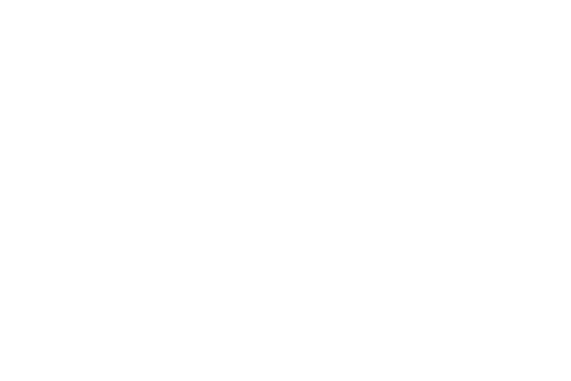}}\\[1ex]
\includegraphics[width=18.5cm]{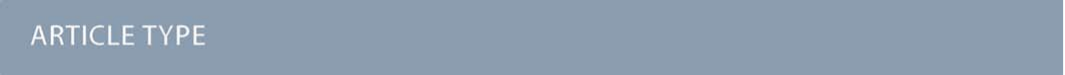}}\par
\vspace{1em}
\sffamily
\begin{tabular}{m{4.5cm} p{13.5cm} }

\includegraphics{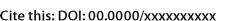} & \noindent\LARGE{\textbf{A Python workflow definition for computational materials design$^\dag$}} \\
\vspace{0.3cm} & \vspace{0.3cm} \\

 & \noindent\large{Jan Janssen,$^{\ast}$\textit{$^{a}$} Janine George,\textit{$^{b, c}$} Julian Geiger,\textit{$^{d}$} Marnik Bercx,\textit{$^{d}$} Xing Wang,\textit{$^{d}$} Christina Ertural,\textit{$^{b}$} Joerg Schaarschmidt,\textit{$^{e}$} Alex M. Ganose,\textit{$^{f}$} Giovanni Pizzi,\textit{$^{d}$} Tilmann Hickel,\textit{$^{a, b}$} and Joerg Neugebauer\textit{$^{a}$}} \\

\includegraphics{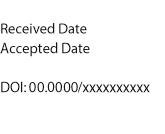} & \noindent\normalsize{Numerous Workflow Management Systems (WfMS) have been developed in the field of computational materials science with different workflow formats, hindering interoperability and reproducibility of workflows in the field. To address this challenge, we introduce here the Python Workflow Definition (PWD) as 
a workflow exchange format to share workflows between Python-based WfMS, currently AiiDA, jobflow, and pyiron.
This development is motivated by the similarity of these three Python-based WfMS, that represent the different workflow steps and data transferred between them as nodes and edges in a graph. With the PWD, we aim at fostering the interoperability and reproducibility between the different WfMS in the context of Findable, Accessible, Interoperable, Reusable (FAIR) workflows.
To separate the scientific from the technical complexity, the PWD consists of three components: (1) a conda environment that specifies the software dependencies, (2) a Python module that contains the Python functions represented as nodes in the workflow graph, and (3) a workflow graph stored in the JavaScript Object Notation (JSON).
The first version of the PWD supports directed acyclic graph (DAG)-based workflows. Thus, any DAG-based workflow defined in one of the three WfMS can be exported to the PWD and afterwards imported from the PWD to one of the other WfMS. After the import, the input parameters of the workflow can be adjusted and computing resources can be assigned to the workflow, before it is executed with the selected WfMS. This import from and export to the PWD is enabled by the PWD Python library that implements the PWD in AiiDA, jobflow, and pyiron.} \\

\end{tabular}

 \end{@twocolumnfalse} \vspace{0.6cm}

  ]

\renewcommand*\rmdefault{bch}\normalfont\upshape
\rmfamily
\section*{}
\vspace{-1cm}


\footnotetext{\textit{$^{a}$~Max Planck Institute for Sustainable Materials, 40237 D\"usseldorf, Germany}}
\footnotetext{\textit{$^{b}$~Bundesanstalt f\"ur Materialforschung und -pr\"ufung, 12205 Berlin, Germany}}
\footnotetext{\textit{$^{c}$~Friedrich-Schiller Universit\"at Jena, 07743 Jena, Germany}}
\footnotetext{\textit{$^{d}$~PSI Center for Scientific Computing, Theory and Data, 5232 Villigen PSI, Switzerland}}
\footnotetext{\textit{$^{e}$~Karlsruhe Institute of Technology (KIT), 76344 Eggenstein-Leopoldshafen, Germany}}
\footnotetext{\textit{$^{f}$~Imperial College London, 80 Wood Lane, W12 7TA London, UK}}

\footnotetext{\dag~Supplementary Information available: [details of any supplementary information available should be included here]. See DOI: 00.0000/00000000.}

\footnotetext{$\ast$~janssen@mpi-susmat.de}


\section{Introduction}
Due to their intrinsic hierarchical nature, material properties depend on the coupling of various domains, among others, materials chemistry, defect engineering, microstructure physics, and mechanical engineering.
This often requires multiscale simulation approaches to adequately model materials with different communities representing the different scales.
Consequently, the goal of multiscale simulations in materials science is to bridge the gap between the macroscale relevant for applying these materials and the quantum mechanical \textit{ab initio} approach of a universal parameter-free description of materials at the atomic scale. 
One of these multiscale simulation approaches that has recently gained popularity is coupling the electronic-structure scale and atomic scale by training machine-learned interatomic potentials (MLIP)~\cite{mlpot}. 
Such a training of a MLIP typically consists of the generation of a reference dataset of electronic structure simulations, the fitting of the MLIP with a specialized fitting code, typically written in Python based on machine learning frameworks like pytorch and tensorflow, and the validation of the MLIP with atomistic simulations, often with widespread software such as the Large-scale Atomic/Molecular Massively Parallel Simulator (LAMMPS)~\cite{lammps} or the atomic simulation environment (ASE)~\cite{ase}, both of which also provide Python interfaces.
Consequently, it requires expertise in electronic structure simulations, in fitting the MLIP, as well as in interatomic potential simulation, with the corresponding simulation and fitting codes being developed by different communities~\cite{Menon2024}. 
The resulting challenge of managing simulation codes from different communities in a combined study of hundreds or thousands of simulations has led to the development of a number of Workflow Management Systems (WfMS). Similarly, high-throughput screening studies, which also couple large numbers of simulations executed with simulation codes at different scales, with different computational costs, and developed from different communities, benefit from WfMS. 

In this context, a scientific workflow is commonly defined as the reproducible protocol of a series of process steps, including the transfer of information between them~\cite{battery2030, bekemeier}. This can be visualized as a graph with the nodes referencing the computational tools and the edges the information transferred between those nodes. Correspondingly, a WfMS is a software tool to orchestrate the construction, management, and execution of the workflow~\cite{fairworkflows}. The advantages of using a WfMS are: (1) Automized execution of the workflow nodes on high-performance computing (HPC) clusters; (2) improved reproducibility, documentation, and distribution of workflows based on a standardized format; (3) user-friendly interface for creating, editing, and executing workflows; (4) interoperability of scientific software codes; (5) orchestration of high-throughput studies with a large number of individual calculations; (6) out-of-process caching of the data transferred via the edges of the workflow and storage of the final results; (7) interfaces to community databases for accessing and publishing data~\cite{bekemeier}.  
As a consequence, using a WfMS abstracts the technical complexity, and the workflow centers around the scientific complexity.

In contrast to WfMS in other communities like BioPipe~\cite{biopipe}, which defines workflows in the Extensible Markup Language (XML), or SnakeMake~\cite{snakemake}, NextFlow~\cite{nextflow} and Common Workflow Language~\cite{cwl}, which introduce their own workflow languages, many WfMS in the computational materials science community use Python as the workflow language~\cite{aiida, aiida1, asr, jobflow, psiflow, myqueue, wfl, pyiron, parsl, signac, perqueue, qmflows}.
Using a programming language to define workflows has the benefit that flow control elements, like loops and conditionals, are readily available as basic features of the language, which is not the case for static languages.
This is a limitation of static languages, such as XML (more on this in Sec.~1 and the supporting information).
Furthermore, the choice of Python in the field of computational materials science has three additional advantages:
(1) the Python programming language is easy to learn as its syntax is characterized by very few rules and special cases, resulting in better readability compared to most workflow languages and a large number of users in the scientific community,  
(2) the improved computational efficiency of transferring large amounts of small data objects between the different workflow steps in-memory, compared to file-based input and output (IO), and 
(3) a large number of scientific libraries for the Python programming language, including many for machine learning, materials science and related domain sciences.

The increasing number of WfMS being developed in the computational materials science community and beyond led to the development of benchmarks implementing the same workflow in different WfMS~\cite{nfdi4ing} and the extension of the FAIR (Findable, Accessible, Interoperable, and Reusable) principles to FAIR workflows~\cite{fairworkflows}. However, the interoperability between different WfMS remains challenging, even within the subgroup of WfMS that use Python as the workflow language. For this specific case, three levels of interoperability can be identified:
(1)~the same scientific Python functions are shared between multiple WfMS, e.g., parsers for the input and output files of a given simulation code,
(2)~the Python functions representing the nodes and the corresponding edges are shared as a template, so that the same workflow can be executed with multiple WfMS and
(3)~the workflow template, including the intermediate results of the workflow, e.g., the inputs and outputs of each node, is shared.

In the following, the Python Workflow Definition (PWD) for directed acyclic graphs (DAG) and the corresponding Python interface~\cite{repository} are introduced. They implement the second level of interoperability for the following three WfMS: AiiDA~\cite{aiida, aiidaworkflows, aiida1}, jobflow~\cite{jobflow}, and pyiron~\cite{pyiron}. The interoperability of the PWD is demonstrated in three examples: (1)~The coupling of Python functions, (2)~the calculation of an energy-versus-volume curve with the Quantum ESPRESSO Density Functional Theory (DFT) simulation code~\cite{quantumespresso, quantumespresso2} and~(3) the benchmark file-based workflow for a finite element simulation introduced in Ref.~\cite{nfdi4ing}. These three examples highlight the application of the PWD to pure Python workflows, file-based workflows based on calling external executables with file transfer between them, and mixed workflows that combine Python functions and external executables.

\begin{figure}[t]
    \centering
    \includegraphics[width=0.45\textwidth]{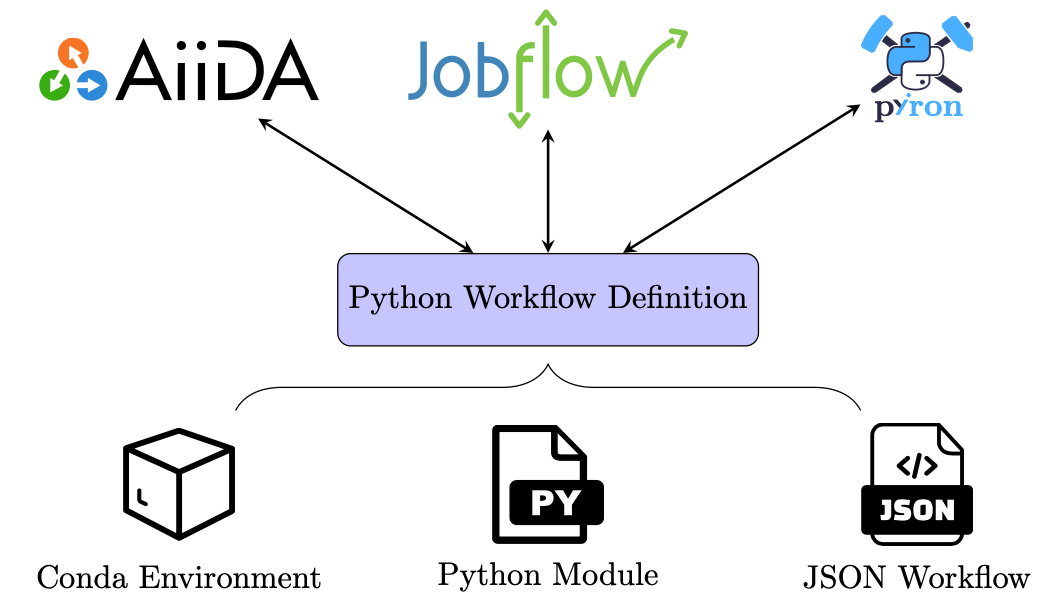}
    \caption{The Python Workflow Definition (PWD) consists of three components: a conda environment, a Python module, and a JSON workflow representation. The three Workflow Management Systems AiiDA, jobflow, and pyiron all support both importing and exporting to and from the PWD.}
    \label{fig:pwd}
\end{figure}%

\section{Python Workflow Definition}
Following the goal of separating technical complexity from  scientific complexity, our suggestion for a PWD consists of three parts: 
(1)~The software dependencies of the workflow are specified in a conda environment file, so all dependencies can be installed using the conda package manager, which is commonly used in the scientific community~\cite{bioconda}.
(2)~Additional Python functions, which represent the nodes in the workflow graph, are provided in a separate Python module.
(3)~Finally, the workflow graph with nodes and edges is stored in the JavaScript Object Notation (JSON) with the nomenclature inspired by the Eclipse Layout Kernel (ELK) JSON format~\cite{elk}.
This is illustrated in Fig.~\ref{fig:pwd}, together with the three WfMS currently supporting the PWD.
If all the involved scientific functionalities are already available within preexisting conda packages, the Python module (part 2) is not required. Still, while an increasing number of open-source simulation codes and utilities for atomistic simulations are available on conda for different scientific domains~\cite{bioconda}, in most cases, additional Python functions are required. These functions are typically stored in the Python module. 

As a first simple example workflow, the addition of the product and quotient of two numbers, $c = a/b + a \cdot b$, and subsequent squaring of their sum is represented in the PWD.
To illustrate the coupling of multiple Python functions, this computation is split into three Python functions, 
a \mintinline{python}{get_prod_and_div()} function to compute the product and quotient of two numbers,
a \mintinline{python}{get_sum()} function for the summation, 
and a \mintinline{python}{get_square()} function to raise the number to the power of two:

\begin{minted}[xleftmargin=20pt]{python}
def get_prod_and_div(
    x: float = 1.0, y: float = 1.0
) -> dict[str, float]:
    return {"prod": x * y, "div": x / y}

def get_sum(x, y):
    return x + y

def get_square(x):
    return x**2
\end{minted}

It is important to note here, that the Python functions are defined independently of a specific WfMS, so they can be reused with any WfMS or even without. Furthermore, the Python functions highlight different levels of complexity supported by the PWD:
The \mintinline{python}{get_prod_and_div()} function returns a dictionary with two output variables, with the keys \mintinline{python}{"prod"} and \mintinline{python}{"div"} referencing the product and quotient of the two input parameters.
Instead, the summation function \mintinline{python}{get_sum()} takes two input variables and returns only a single output, which is then fed into the \mintinline{python}{get_square()} function that returns the final result.
In addition, the \mintinline{python}{get_prod_and_div()} function uses default parameter values and type hints, which are optional features of the Python programming language supported by the PWD to improve the interoperability of the workflow.
While the computation of the product and quotient of two numbers could be done in two separate functions, the purpose here is to demonstrate the implementation of a function with more than one return value.
Another example of such a function could be a matrix diagonalization function that returns the eigenvalues and eigenvectors.
The supplementary information provides a more in-depth discussion of how function returns are resolved to an unambiguous mapping in the graph.

As a demonstration, the Python functions 
\mintinline{python}{get_prod_and_div()}, \mintinline{python}{get_sum()} and \mintinline{python}{get_square()} are stored in a Python module named \mintinline{python}{workflow.py}. In addition, as these functions have no dependencies other than the Python standard library, the conda environment, \mintinline{python}{environment.yml}, is sufficiently defined by specifying the Python version:

\begin{minted}[xleftmargin=20pt]{yaml}
channels:
- conda-forge
dependencies:
- python=3.12
\end{minted}

The conda-forge community channel is selected as the package source as it is freely available and provides a large number of software packages for materials science and related disciplines~\cite{bioconda}.
For other examples, e.g., the calculation of the energy-versus-volume curve with Quantum ESPRESSO (see below), the conda environment would contain the software dependencies of the workflow, including the simulation code and additional utilities like parsers.
It is important to note that the combination of the Python module and the conda environment already addresses the requirements for the first level of interoperability defined above. As the scientific Python functions are defined independently of any workflow environment, they can be used with any WfMS that supports Python functions as nodes. 

\begin{figure}[t]
    \centering
    \includegraphics[width=0.35\textwidth]{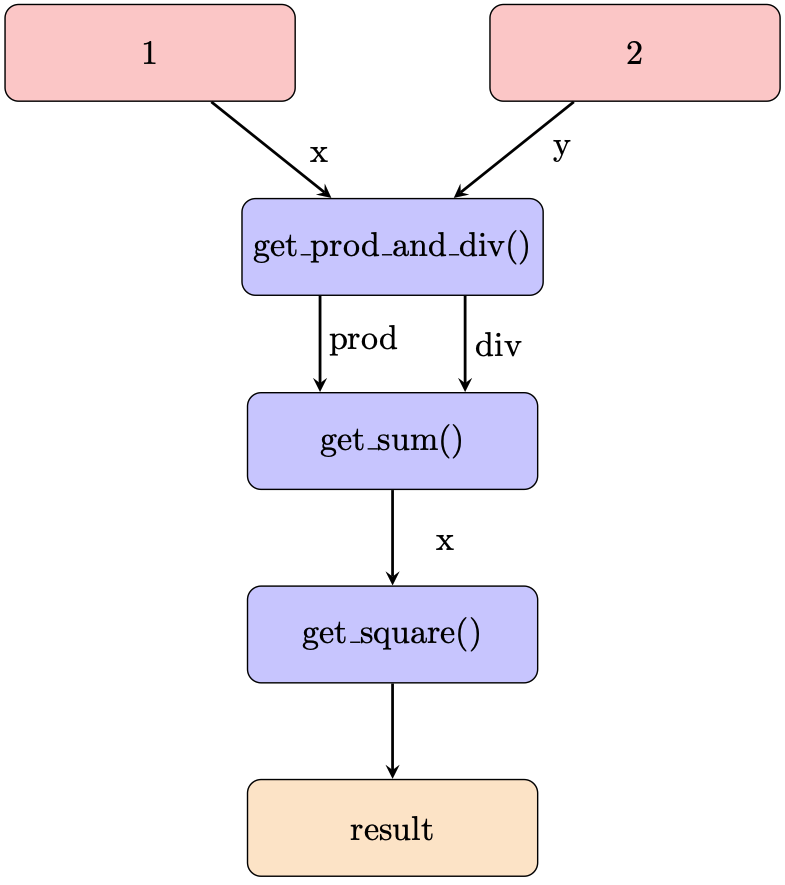}
    \caption{The arithmetic workflow computes the sum of the product and quotient of two numbers. The red nodes of the workflow graph denote inputs, the orange the outputs, and the blue nodes the Python functions for the computations. The labels of the edges denote the data transferred between the nodes.}
    \label{fig:simple}
\end{figure}

The limitation of the first level of interoperability is the loss of connection of the individual functions, that is, which output of one function is reused as input of another function. In terms of the workflow as a graph with the Python functions representing the nodes of the graph, these connections are the edges between the nodes. To define the workflow, we wrap the individual function calls in another function to which we can then pass our input values and from which we retrieve our output value:

\begin{minted}[xleftmargin=20pt]{python}
def workflow(x: float = 1, y: float = 2):
    tmp_dict = get_prod_and_div(x=x, y=y)
    tmp_sum = get_sum(
        x=tmp_dict["prod"], 
        y=tmp_dict["div"],
    )
    return get_square(x=tmp_sum)

result = workflow(x=1, y=2)
\end{minted}

We pass the inputs \mintinline{python}{x=1.0} and \mintinline{python}{y=2.0} to our 
\mintinline{python}{workflow} function, in which the computation of the product and quotient with the \mintinline{python}{get_prod_and_div()} is executed first. This is then followed by a summation of the two results with the \mintinline{python}{get_sum()} function, which returns a single output value that is then fed into the \mintinline{python}{get_square()} function.
The corresponding graph is visualized in Fig.~\ref{fig:simple}.

In the next step, the resulting graph is serialized to an internal JSON representation with the nomenclature and overall structure inspired by the ELK JSON format~\cite{elk}, for sharing the workflow between different WfMS. While human-readable, the JSON format is not intended for direct user interaction, i. e. generating or modifying the JSON with a text editor; rather, it is primarily focused on enabling interoperability of WfMS and long-term storage. For the construction of a workflow, we recommended using one of the existing WfMS and afterwards exporting the workflow to the PWD. The resulting PWD JSON for the arithmetic workflow is:

\begin{minted}[xleftmargin=20pt]{json}
{
  "version": "1.0.0",
  "nodes": [
    {"id": 0, "type": "function", 
     "value": "workflow.get_prod_and_div"},
    {"id": 1, "type": "function", 
     "value": "workflow.get_sum"},
    {"id": 2, "type": "function", 
     "value": "workflow.get_square"},
    {"id": 3, "type": "input",  
     "value": 1, "name": "x"},
    {"id": 4, "type": "input", 
     "value": 2, "name": "y"},
    {"id": 5, "type": "output", 
     "name": "result"}
  ],
  "edges": [
    {"source": 3, "sourcePort": null,
     "target": 0, "targetPort": "x"},
    {"source": 4, "sourcePort": null,
     "target": 0, "targetPort": "y"},
    {"source": 0, "sourcePort": "prod",
     "target": 1, "targetPort": "x"},
    {"source": 0, "sourcePort": "div",
     "target": 1, "targetPort": "y"},
    {"source": 1, "sourcePort": null,
     "target": 2, "targetPort": "x"},
    {"source": 2, "sourcePort": null,
     "target": 5, "targetPort": null}
  ]
}
\end{minted}

On the first level, the PWD JSON format defines the workflow metadata given by the version number, nodes and edges:
\begin{itemize}
\item The version number (of the PWD JSON format) is given by three non-negative integers combined in a string, to enable semantic versioning. Minor changes and patches which do not affect the backwards compatibility are indicated by increasing the second and third numbers, respectively. In contrast, an increase in the first number indicates changes that are no longer backwards compatible.
\item The nodes section is (in this example) a list of six items: The three Python functions defined in the \mintinline{python}{workflow.py} Python module, the two input parameters for the workflow, in this case \mintinline{python}{x=1.0} and \mintinline{python}{y=2.0}, and the output data node. Each node is defined as a dictionary consisting of an \mintinline{python}{"id"}, a \mintinline{python}{"type"}, and a \mintinline{python}{"value"}. In case of the \mintinline{python}{"input"} and \mintinline{python}{"output"} data nodes, the \mintinline{python}{"name"} is an identifier that denotes how the inputs and outputs are exposed by the overall workflow. Moreover, for \mintinline{python}{"input"} data nodes, the \mintinline{python}{"value"} is an optional default value (if provided during workflow construction).
On the other hand, for \mintinline{python}{"function"} nodes, the \mintinline{python}{"value"} entry contains the module and function name.
The usage of the dictionary format allows future extensions by adding additional keys to the dictionary for each node. 
\item In analogy to the nodes, also the edges are stored as a list of dictionaries. The first two edges connect the input parameters with the \mintinline{python}{get_prod_and_div()} function. Each edge is defined based on the source node \mintinline{python}{"source"}, the source port \mintinline{python}{"sourcePort"}, the target node \mintinline{python}{"target"} and the target port \mintinline{python}{"targetPort"}.
As the input data nodes do not have associated ports, their source ports are null. In contrast, the target ports are the input parameters \mintinline{python}{x} and \mintinline{python}{y} of the \mintinline{python}{get_prod_and_div()} function. The PWD JSON representation also contains two edges that connect the two outputs from the \mintinline{python}{get_prod_and_div()} function to the inputs of the \mintinline{python}{get_sum()} function. In analogy to the target port, the source port specifies the output dictionary key to select from the output. If no source port is available (typically because a function does not return a dictionary containing keys that can serve as source ports), then the source port is set to \mintinline{python}{null} and, in that case, the entire return value of the function (possibly, also a tuple, list, dictionary or any other Python data type) is transferred to the target node. This is the case for the fifth edge that maps the return value of the \mintinline{python}{get_sum()} function to the \mintinline{python}{"x"} input of the \mintinline{python}{get_square()} function. Finally, its result is exposed as the global \mintinline{python}{"result"} output of the workflow, the last edge in the graph. As the \mintinline{python}{get_square()} function does return the value directly, and the target of the edge is an output data node (that does not define a port), both \mintinline{python}{"targetPort"} and \mintinline{python}{"sourcePort"} are null in this edge.
\end{itemize}

By using a list of dictionaries for both the nodes and edges, as well as a dictionary at the first level, the PWD JSON format is extensible, and additional metadata beyond the version number can be added in the future. As the focus of this first version of the PWD is the interoperability between the different WfMS, apart from the node types (useful for parsing and validation), no additional metadata is included in the PWD JSON format. To assist the users in analyzing the JSON representation of the PWD, the PWD Python interface provides a \mintinline{python}{plot()} function to visualize the workflow graph. The \mintinline{python}{plot()} function is introduced in the supplementary material.

\section{Implementation} 
The focus of the PWD is to enable the interoperability between different WfMS. Thus, it is recommended that users always use one of the supported WfMS to create the workflow and export it to the PWD using the PWD Python library. Afterwards, the workflow can be imported into a different WfMS, the input parameters can be modified, and computational resources can be assigned before the workflow is executed. In the following, the same workflow introduced above is defined in AiiDA, jobflow, and pyiron. This highlights the similarities between these Python-based WfMS, which all use the Python programming language as their workflow language, with the selection of WfMS being based on the authors' experience. 
While this section covers the export of the workflow to the WfMS, the import is discussed in the application section below. Finally, interfaces for additional WfMS are planned in the future. Full integration will be achieved with PWD support becoming an integral part of the WfMS itself and the PWD package possibly becoming a dependency.

\subsection{AiiDA}
The ``Automated Interactive Infrastructure and Database for Computational Science'' (AiiDA)~\cite{aiida, aiidaworkflows, aiida1} is a WfMS with a strong focus on data provenance and high-throughput performance.
AiiDA provides checkpointing, caching, and error handling features for dynamic workflows at full data provenance (via an SQL database), among other features.
While it originated from the field of computational materials science\cite{Huber2021}, it has recently been extended to several other fields (see e.g. the codes supported in the AiiDA plugin registry\cite{aiidapluginregistry}) and to experiments\cite{Kraus24}.
In the following code snippets, we will be using the \mintinline{python}{WorkGraph}, a recently added and actively developed new AiiDA workflow component\cite{workgraph-rtd}.
The \mintinline{python}{WorkGraph} functions like a canvas for workflow creation to which a user can dynamically add \mintinline{python}{Tasks}, that is, workflow components (also called ``nodes'' in a graph-based representation of a workflow), and connect them with \mintinline{python}{Links} (the ``edges`` in the PWD).
This approach to workflow creation offers the flexibility of dynamically chaining workflow components together ``on-the-fly'', an approach especially crucial for rapid prototyping common in scientific environments.
Implementation of the arithmetic workflow is shown in the following snippets. It starts with the import of relevant modules:
\begin{minted}[xleftmargin=20pt]{python}
import python_workflow_definition as pwd

from aiida import orm, load_profile
from aiida_workgraph import WorkGraph, task

from arithmetic_workflow import (
    get_sum as _get_sum,
    get_prod_and_div as _get_prod_and_div,
    get_square as _get_square
)

load_profile()
\end{minted}

We first import the \mintinline{python}{python_workflow_definition} module, which contains the necessary code to import from and export to the general Python workflow definition. In addition, from the AiiDA core module, we import AiiDA's Object-Relational Mapper (ORM), as well as the \mintinline{python}{load_profile} function. The ORM module allows mapping Python data types to the corresponding entries in AiiDA's underlying SQL database, and calling the \mintinline{python}{load_profile} function ensures that an AiiDA profile (necessary for running workflows via AiiDA) is loaded. From the \mintinline{python}{aiida-workgraph} module, we import the main \mintinline{python}{WorkGraph} class, as well as the \mintinline{python}{task} decorator. Lastly, we import the Python functions from the \mintinline{python}{arithmetic_workflow} module.

To convert the pure Python functions from the arithmetic workflow into AiiDA WorkGraph workflow components, we wrap them with the \mintinline{python}{task} function (decorator):
\begin{minted}[xleftmargin=20pt]{python}
get_prod_and_div = task(outputs=["prod", "div"])(
    _get_prod_and_div
)
get_sum = task()(_get_sum)
get_square = task()(_get_square)
\end{minted}

As the \mintinline{python}{get_prod_and_div} function returns a dictionary with multiple outputs, we pass this information to the \mintinline{python}{task} function via the \mintinline{python}{outputs} argument, such that we can reference them at a later stage (they will become the ports in the PWD JSON). Without the \mintinline{python}{outputs} argument, the whole output dictionary \mintinline{python}{{"prod": x * y, "div": x / y}} would be wrapped as one port with the default \mintinline{python}{"result"} key. This is what actually happens to the single return value of the \mintinline{python}{get_sum()} function (as further outlined in the supplementary information, we follow a similar approach to resolve the ``ports'' entries in the ``edges'' of the PWD).
Next follows the instantiation of the WorkGraph:
\begin{minted}[xleftmargin=20pt]{python}
wg = WorkGraph("arithmetic")
\end{minted}

Which then allows adding the previously defined \mintinline{python}{Tasks}:
\begin{minted}[xleftmargin=20pt]{python}
get_prod_and_div_task = wg.add_task(
    get_prod_and_div,
    x=orm.Float(1.0),
    y=orm.Float(2.0),
)
get_sum_task = wg.add_task(
    get_sum,
    x=get_prod_and_div_task.outputs.prod,
    y=get_prod_and_div_task.outputs.div,
)
get_square_task = wg.add_task(
    get_square,
    x=get_sum_task.outputs.result,
)
\end{minted}
Here, we wrap the inputs as AiiDA ORM nodes to ensure they are registered as nodes when exporting to the PWD.
Further, in the \mintinline{python}{get_sum_task}, the outputs of the previous \mintinline{python}{get_prod_and_div_task} are passed as inputs. Note that at this stage, the workflow has not been run, and these output values do not exist yet. In WorkGraph, such outputs are represented by a \mintinline{python}{Socket} that serves as a placeholder for future values and already allows linking them to each other in the workflow:
\begin{minted}[xleftmargin=20pt]{python}
In [1]: print(get_prod_and_div_task.outputs.prod)
Out[1]: SocketAny(name="prod", value=None)
\end{minted}

Alternatively, adding tasks to the WorkGraph and linking their outputs can also be done in two separate steps, shown below for linking the \mintinline{python}{get_prod_and_div_task} and \mintinline{python}{get_sum_task}:
\begin{minted}[xleftmargin=20pt]{python}
...
get_sum_task = wg.add_task(
    get_sum,
)
wg.add_link(
    get_prod_and_div_task.outputs.prod,
    get_sum_task.inputs.x,
)
wg.add_link(
    get_prod_and_div_task.outputs.div,
    get_sum_task.inputs.y,
)
...
\end{minted}
Lastly, the JSON file containing the PWD can be written to disk via:
\begin{minted}[xleftmargin=20pt]{python}
    pwd.aiida.write_workflow_json(
        wg=wg,
        file_name="arithmetic.json"
    )
\end{minted}

\subsection{jobflow}
Jobflow~\cite{jobflow} was developed to simplify the development of high-throughput workflows. It uses a decorator-based approach to define the \mintinline{python}{Job}'s that can be connected to form complex workflows (\mintinline{python}{Flow}s). Jobflow is the workflow language of the workflow library atomate2\cite{ganose_atomate2_2025}, designed to replace atomate\cite{mathew_atomate_2017}, which was central to the development of the Materials Project~\cite{materialsproject} database.

First, the \mintinline{python}{job} decorator, which allows the creation of \mintinline{python}{Job} objects, and the \mintinline{python}{Flow} class are imported. In addition, the PWD Python module and the functions of the arithmetic workflow are imported in analogy to the previous example.

\begin{minted}[xleftmargin=20pt]{python}
from jobflow import job, Flow
import python_workflow_definition as pwd
from arithmetic_workflow import (
    get_sum as _get_sum, 
    get_prod_and_div as _get_prod_and_div,
    get_square as _get_square,
)
\end{minted}

Using the job object decorator, the imported functions from the arithmetic workflow are transformed into jobflow \mintinline{python}{Job}s. These \mintinline{python}{Job}s can delay the execution of Python functions and can be chained into workflows (\mintinline{python}{Flow}s). A \mintinline{python}{Job} can return serializable outputs (e.g., a number, a dictionary, or a Pydantic model) or a so-called \mintinline{python}{Response} object, which enables the execution of dynamic workflows where the number of nodes is not known prior to the workflow's execution. As jobflow itself is only a workflow language, the workflows are typically executed on high-performance computers with a workflow manager such as Fireworks~\cite{fireworks} or jobflow-remote\cite{jobflow-remote}. For smaller and test workflows, simple linear, non-parallel execution of the workflow graph can be performed with jobflow itself. All outputs of individual jobs are saved in a database. For high-throughput applications, typically, a MongoDB database is used. For testing and smaller workflows, a memory database can be used instead. In Fireworks, its predecessor in the Materials Project infrastructure, this option did not exist, which was a significant drawback.

\begin{minted}[xleftmargin=20pt]{python}
get_prod_and_div = job(_get_prod_and_div)
get_sum = job(_get_sum)
get_square = job(_get_square)

prod_and_div = get_prod_and_div(x=1.0, y=2.0)
tmp_sum = get_sum(
    x=prod_and_div.output.prod,
    y=prod_and_div.output.div,
)
result = get_square(x=tmp_sum.output)

flow = Flow([prod_and_div, tmp_sum, result])     
\end{minted}

As before in the AiiDA example, the workflow has not yet been run. \mintinline{python}{prod_and_div.output.div} 
refers to an \mintinline{python}{OutputReference} object instead of the actual output.

Finally, after the workflow is constructed, it can be exported to the PWD using the PWD Python package to store the jobflow workflow in the JSON format. 

\begin{minted}[xleftmargin=20pt]{python}
pwd.jobflow.write_workflow_json(
    flow=flow, 
    file_name="arithmetic.json",
)    
\end{minted}

\subsection{pyiron}
The pyiron WfMS was developed with a focus on rapid prototyping and up-scaling atomistic simulation workflows~\cite{pyiron}. It has since been extended to support simulation workflows at different scales, including the recent extension to experimental workflows~\cite{pyironexp}. Based on this generalization, the same arithmetic Python workflow is implemented in the pyiron WfMS. Starting with the import of the pyiron job object decorator and the PWD Python module, the functions of the arithmetic workflow are imported in analogy to the previous examples above. 

\begin{minted}[xleftmargin=20pt]{python}
from pyiron_base import job
import python_workflow_definition as pwd
from arithmetic_workflow import (
    get_sum as _get_sum, 
    get_prod_and_div as _get_prod_and_div,
    get_square as _get_square,
)
\end{minted}
    
Using the job object decorator, the imported functions from the arithmetic workflow are converted to pyiron job generators. These job generators can be executed like Python functions; still, internally, they package the Python function and corresponding inputs in a pyiron job object, which enables the execution on HPC clusters by assigning dedicated computing resources and provides the permanent storage of the inputs and output in the Hierarchical Data Format (HDF5). For the \mintinline{python}{get_prod_and_div()} function, an additional list of output parameter names is provided, which enables the coupling of the functions before the execution, to construct the workflow graph.

\begin{minted}[xleftmargin=20pt]{python}
get_sum = job(_get_sum)
get_prod_and_div = job(
    _get_prod_and_div, 
    output_key_lst=["prod", "div"],
)
get_square = job(_get_square)
\end{minted}

After the conversion of the Python functions to pyiron job generators, the workflow is constructed. The pyiron job generators are called just like Python functions; still, they return pyiron delayed job objects rather than the computed values. These delayed job objects are linked with each other by using a delayed job object as an input to another pyiron job generator. Finally, the whole workflow would be only executed once the pull function \mintinline{python}{pull()} is called on the delayed pyiron object of the \mintinline{python}{get_square()} function. At this point, the delayed pyiron objects are converted to pyiron job objects, which are executed using the pyiron WfMS. In particular, the conversion to pyiron job objects enables the automated caching to the hierarchical data format (HDF5) and the assignment of computing resources.

\begin{minted}[xleftmargin=20pt]{python}
prod_and_div = get_prod_and_div(x=1.0, y=2.0)
tmp_sum = get_sum(
    x=prod_and_div.output.prod,
    y=prod_and_div.output.div,
)
result = get_square(x=tmp_sum)
\end{minted}
    
For the example here, the workflow execution is skipped and the workflow is exported to the PWD using the PWD Python package to store the pyiron workflow in JSON format. The export command is implemented in analogy to the export commands for AiiDA and jobflow, taking a delayed pyiron object as an input in combination with the desired file name for the JSON representation of the workflow graph. 

\begin{minted}[xleftmargin=20pt]{python}
pwd.pyiron_base.write_workflow_json(
    delayed_object=result, 
    file_name="arithmetic.json",
)
\end{minted}
    
The implementation of the arithmetic workflow in pyiron demonstrates the similarities to AiiDA and jobflow. 

\begin{figure}[h]
    \centering
    \includegraphics[width=0.45\textwidth]{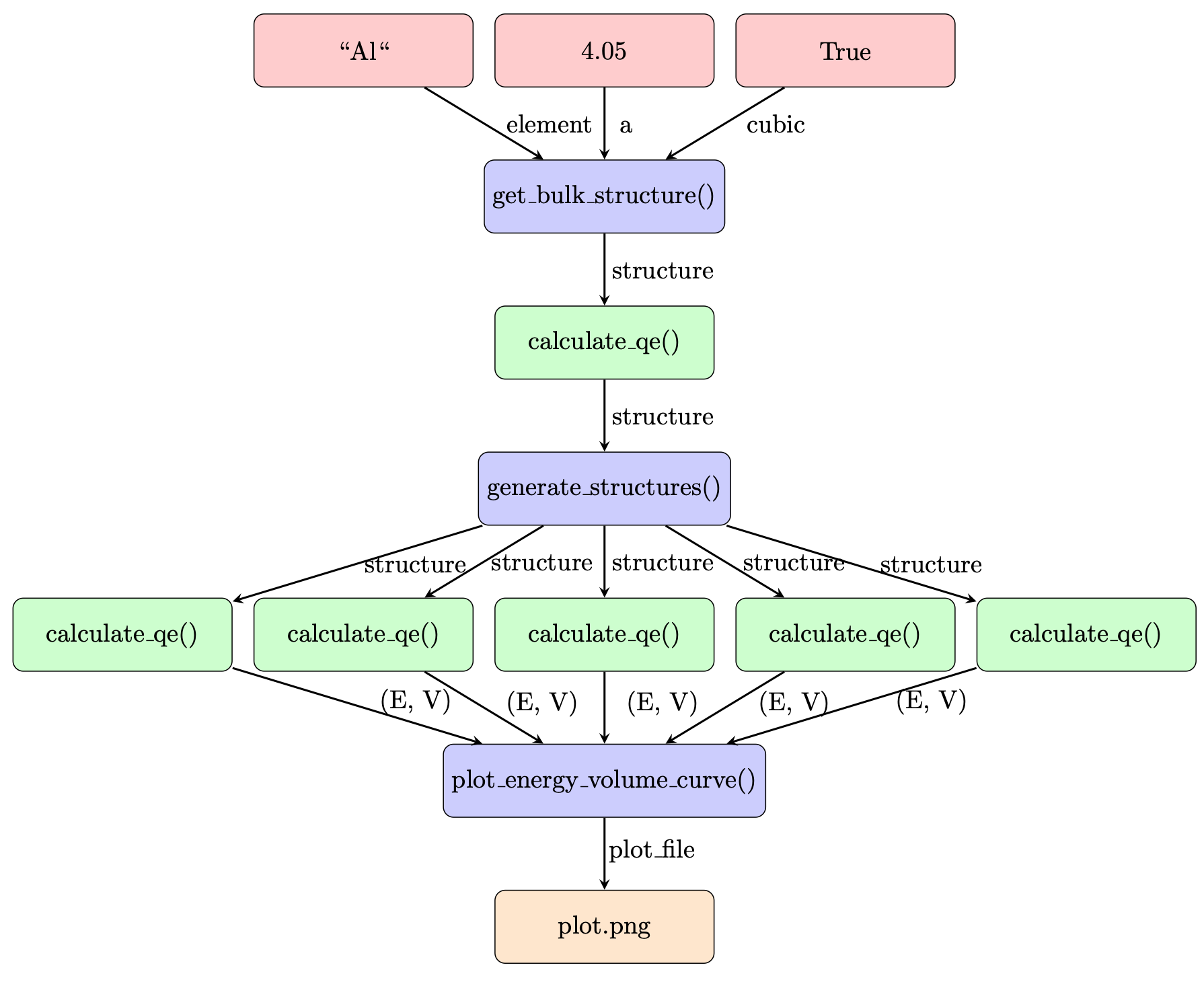}
    \caption{Energy-versus-volume curve calculation workflow with Quantum ESPRESSO. Red boxes denote inputs, orange boxes outputs, blue boxes Python functions and green boxes calls to external executables.}
    \label{fig:evcurve}
\end{figure}%

\section{Application}
To demonstrate the application of the PWD beyond just the arithmetic example above, we consider a second workflow that describes the calculation of an energy-versus-volume curve with Quantum ESPRESSO. The energy-versus-volume curve is typically employed to calculate the equilibrium volume and the compressive bulk modulus for bulk materials. The workflow is illustrated in Fig.~\ref{fig:evcurve}, with the red and orange nodes marking the inputs and outputs of the workflow, the blue nodes the Python functions, and the green nodes indicating Python functions that internally launch Quantum ESPRESSO simulations. The individual steps of the workflow are: 
\begin{enumerate}
\item Based on the input of the chemical element, the lattice constant, and the crystal symmetry, the atomistic bulk structure is generated by calling the bulk structure generation function \mintinline{python}{get_bulk_structure()}. This function is obtained via the Atomistic Simulation Environment (ASE)~\cite{ase} and extended to enable the serialization of the atomistic structure to the JSON format using the OPTIMADE~\cite{optimade} Python tools~\cite{optimadept}.
\item The structure is relaxed afterwards with Quantum ESPRESSO to get an initial guess for the equilibrium lattice constant. Quantum ESPRESSO is written in FORTRAN and does not provide Python bindings, so that the communication is implemented in the \mintinline{python}{calculate_qe()} function by writing input files, calling the external executable, and parsing the output files.  
\item Following the equilibration, the resulting structure is strained in the function \mintinline{python}{generate_structures()} with two compressive strains of -10\% and -5\% and two tensile strains of 5\% and 10\%. Together with the initially equilibrated structure, this leads to a total of five structures.
\item Each structure is again evaluated with Quantum ESPRESSO to compute the energy of the strained structure. 
\item After the evaluation with Quantum ESPRESSO, the calculated energy-volume pairs are collected in the \mintinline{python}{plot_energy_volume_curve()} function and plotted as an energy-versus-volume plot. The final plot is saved in a file named \mintinline{python}{plot.png}.
\end{enumerate}

Compared to the previous arithmetic example, this workflow is more advanced and not only illustrates one-to-one connections, in terms of one node being connected to another node, but also one-to-many and many-to-one connections. The latter two are crucial to construct the loop over different strains, compute the corresponding volume and energy pairs, and gather the results in two lists, one for the volumes and one for the energies, to simplify plotting. In addition, it highlights the challenge of workflows in computational materials science to couple Python functions for structure generation, modifications, and data aggregation with simulation codes that do not provide Python bindings and require file-based communication. Given the increased complexity of the workflow, the implementation for the individual WfMS is provided in the supplementary material. Instead, the following briefly highlights how the workflow, which was previously stored in the PWD, can be reloaded with the individual frameworks. 

Starting with the AiiDA WfMS, the first step is to load the \mbox{AiiDA} profile and import the PWD Python interface. Afterwards, the workflow can be loaded from the JSON representation \mintinline{python}{qe.json} using the \mintinline{python}{load_workflow_json()} function.
To demonstrate the capability of modifying the workflow parameters before the execution of the (re-)loaded workflow, we then modify the lattice constant of the \mintinline{python}{get_bulk_structure()} node to {4.05\AA}. Similarly, one could also adapt the element, bulk structure, or strain list input parameters of the workflow.
Finally, the workflow is executed by calling the \mintinline{python}{run()} function of the AiiDA WorkGraph object: 

\begin{minted}[xleftmargin=20pt]{python}
from aiida import orm, load_profile   
import python_workflow_definition as pwd

load_profile()

wg = pwd.aiida.load_workflow_json(
    file_name="qe.json"
)
wg.tasks[0].inputs.a.value = orm.Float(4.05)
wg.run()
\end{minted}

The same JSON representation \mintinline{python}{qe.json} of the workflow can also be loaded with the jobflow WfMS. Again, the jobflow WfMS and the PWD Python interface are imported. The JSON representation \mintinline{python}{qe.json} is loaded with the \mintinline{python}{load_workflow_json()} function. Afterwards, the lattice constant is adjusted to {4.05\AA} and finally the workflow is executed with the jobflow \mintinline{python}{run_locally()} function. We note that the same workflow could also be submitted to a HPC cluster, but local execution is primarily chosen here for demonstration purposes to enable the local execution of the provided code examples.

\begin{minted}[xleftmargin=20pt]{python}
from jobflow.managers.local import run_locally
import python_workflow_definition as pwd

flow = pwd.jobflow.load_workflow_json(
    file_name="qe.json"
)
flow[0].function_kwargs["a"] = 4.05
run_locally(flow)
\end{minted}
    
In analogy to the AiiDA WfMS and the jobflow WfMS. the energy-versus-volume curve workflow can also be executed with the pyiron WfMS. Starting with the import of the PWD Python interface, the JSON representation \mintinline{python}{qe.json} of the workflow is again loaded with the \mintinline{python}{load_workflow_json()} function, followed by the adjustment of the lattice constant to {4.05\AA} by accessing the input of the first delayed job object. Finally, the last delayed job object's \mintinline{python}{pull()} function is called to execute the workflow. 

\begin{minted}[xleftmargin=20pt]{python}
import python_workflow_definition as pwd

wf = pwd.pyiron_base.load_workflow_json(
    file_name="qe.json"
)
wf[0].input["a"] = 4.05
wf[-1].pull()
\end{minted}
    
The focus of this second example is to highlight that a workflow stored in the PWD can be executed with all three workflow frameworks with minimally adjusted code. This not only applies to simple workflows consisting of multiple Python functions but also includes more complex logical structures like the one-to-many and many-to-one connections, covering any Directed Acyclic Graphs (DAG) topology. We remark, though, that in the current version the restriction to DAGs is also a  limitation of the PWD, as it does not cover dynamic workflows, such as a while loop that adds additional steps until a given condition is fulfilled. Another challenge is the assignment of computational resources, like the assignment of a fixed number of CPU cores, as the wide variety of different HPC clusters with different availability of computing resources hinders standardization. As such, the user is required to adjust the computational resources via the WfMS after reloading the workflow graph. For this reason, the workflow is also not directly executed by the \mintinline{python}{load_workflow_json()} function, but rather the user can explore and modify the workflow and afterwards initiate the execution with any of the WfMS once the required computational resources are assigned.   

\begin{figure}[t]
    \centering
    \includegraphics[width=0.95\linewidth]{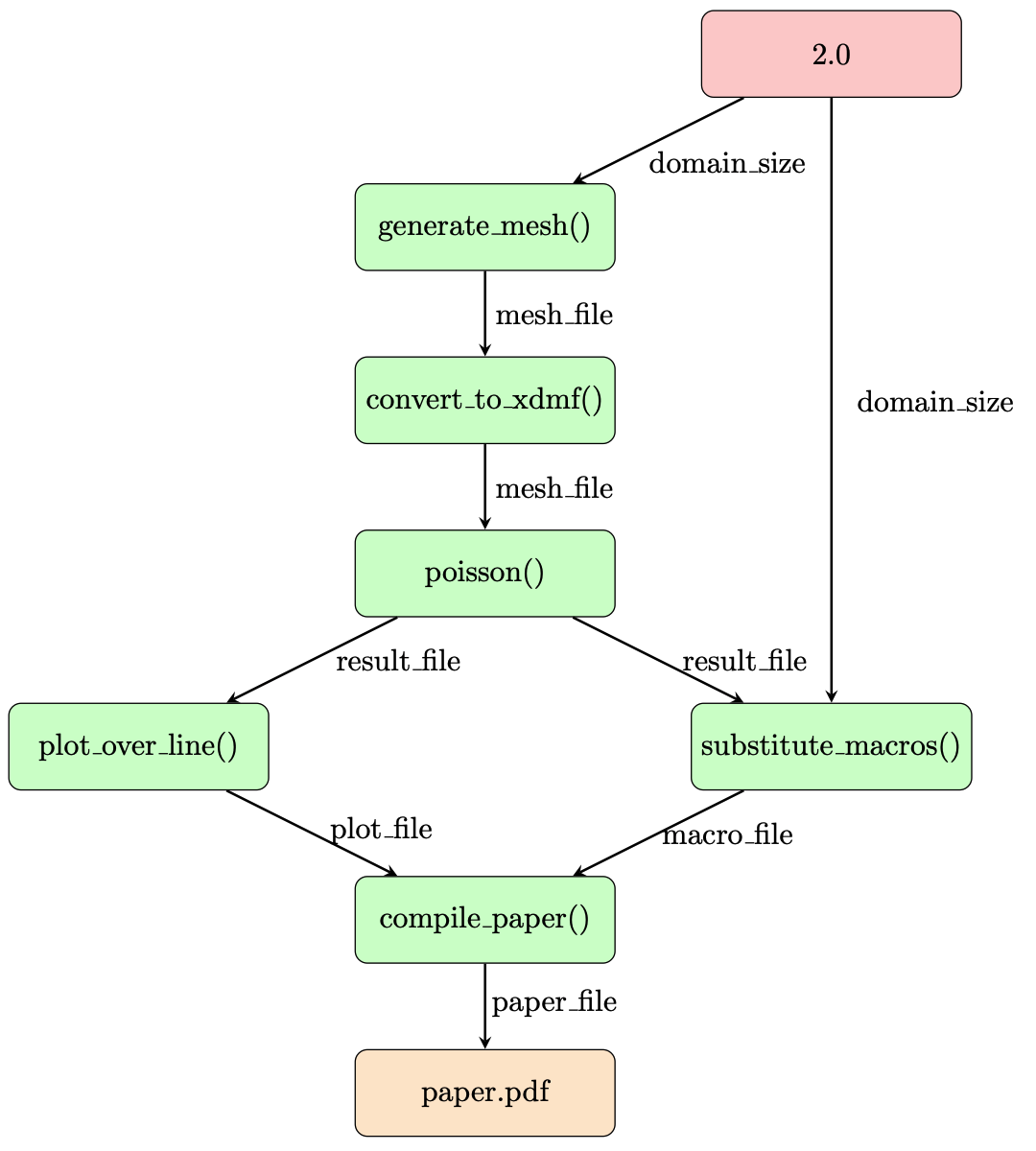}
    \caption{File-based finite element workflow from Ref.~\cite{nfdi4ing} implemented with the Python Workflow Definition (PWD). Red nodes denote inputs, orange nodes outputs, green nodes calls to external executables, and the labels on the edges the files and data transferred between them. Files are passed as path objects between the individual steps.}
    \label{fig:nfdi}
\end{figure}%

\section{Compatibility to non-Python-based workflows}
The two previous examples demonstrated Python-based workflows, which couple either solely Python functions or Python functions and external executables, wrapped by other Python functions that write the input files and parse the output files. Before Python-based WfMS, a number of previous WfMS were introduced, which couple simulation codes solely based on transferring files between the different steps of the workflow~\cite{biopipe, snakemake, nextflow, cwl}. To demonstrate that the PWD can also be applied to these file-based workflows, we implement the benchmark published in Ref.~\cite{nfdi4ing} for file-based workflows in materials science in the PWD. The corresponding workflow is illustrated in Fig.~\ref{fig:nfdi}.

As the file-based workflow for finite element simulations is already discussed in the corresponding publication~\cite{nfdi4ing}, it is only summarized here. A mesh is generated in the first pre-processing step, followed by the conversion of the mesh format in the second pre-processing step. Afterwards, the Poisson solver of the finite element code is invoked. Finally, in the postprocessing, the data is first visualized in a line plot, a TeX macro is generated, and a TeX document is compiled, resulting in the \mintinline{python}{paper.pdf} as the final output. 
To represent this file-based workflow in the PWD, each node is represented by a Python function. This Python function acts as an interface to the corresponding command line tool, handling the writing of the input files, calling of the command line tool and the parsing of the output files. In this specific case, which is purely based on external executables, the output files of one node are copied to be used as input files for the next node, and only the path to the corresponding file is transferred in Python. The Python function for the \mintinline{python}{generate_mesh()} node is given below:
\begin{minted}[xleftmargin=20pt]{python}
import os
from conda_subprocess import check_output
import shutil

def generate_mesh(
    domain_size: float, 
    source_directory: str
) -> str:
    stage_name = "preprocessing"
    output_file_name = "square.msh"
    source_file_name = "unit_square.geo"
    os.makedirs(stage_name, exist_ok=True)
    source_file = os.path.join(
        source_directory, source_file_name
    )
    shutil.copyfile(
        source_file, 
        os.path.join(stage_name, source_file_name)
    )
    _ = check_output(
        [
            "gmsh", "-2", "-setnumber", 
            "domain_size", str(domain_size),
            source_file_name, 
            "-o", output_file_name
        ],
        prefix_name=stage_name,
        cwd=stage_name,
        universal_newlines=True,
    )
    return os.path.abspath(
        os.path.join(stage_name, output_file_name)
    )
\end{minted}
The input parameters of the \mintinline{python}{generate_mesh()} function are the \mintinline{python}{domain_size} and the \mintinline{python}{source_directory} with the \mintinline{python}{source_directory} referencing the location of additional input files. Following the definition of a number of variables, a directory is created and the source files are copied as templates to this directory. Then the external executable is called. Here we use the \mintinline{python}{conda_subprocess} package~\cite{condasubprocess}, which allows us to execute the external executable in a separate conda environment. This was a requirement of the file-based benchmark workflow~\cite{nfdi4ing}. Finally, the path to the output file \mintinline{python}{"square.msh"} is returned as result of the Python function.

While the definition of a Python function for each node is an additional overhead, it is important to emphasize that the Python functions were only defined once, independently of the different WfMS and afterwards the same Python functions were used in all three WfMS. Again, the step-by-step implementation in the three different WfMS and the exporting to the PWD is available in the supplementary material. 
This third example again highlights the universal applicability of the PWD, as it can cover both Python-based workflows and file-based workflows.

Finally, to increase the impact of the PWD and extend its generality beyond the three WfMS discussed in this work, we provide a first proof-of-concept implementation to convert a PWD JSON file to the Common Workflow Language~\cite{cwl}.
In this case each input and output of every node is serialized using the built-in pickle serialization of the Python Standard library.
The resulting pickle files are then transferred from one node to another through CWL. To convert a given PWD JSON file, use the \mintinline{python}{write_workflow()} from the CWL submodule of the PWD Python interface:
\begin{minted}[xleftmargin=20pt]{python}
import python_workflow_definition as pwd

pwd.cwl.write_workflow(
    file_name="workflow.json"
)
\end{minted}

This Python function creates the corresponding CWL files to represent the individual nodes, as well as the resulting workflow in the CWL, which can then be executed by any CWL engine (given that the necessary dependencies are available on the system).
Still, it is important to emphasize that in contrast to the interfaces to the Python-based WfMS, the interface to the CWL is a one-way conversion only from the PWD to the CWL, not the other way around. Furthermore, by converting the workflow to the CWL, the performance benefit of handling the data on the edges of the workflow inside the Python process is lost as the CWL interface is based on file-based communication. Lastly, another notable concept close to the PWD is the graph-based Abstract Syntax Tree (AST)~\cite{ast} representation of the Python standard library. For brevity this comparison is discussed in the supplementary information. 

\section*{Conclusions} 
The Python Workflow Definition (PWD) enables users to develop interoperable workflows to fulfill the requirements for Findable, Accessible, Interoperable and Reusable (FAIR) workflows. The first version of the PWD currently supports Directed Acyclic Graphs (DAGs) based workflows and interoperability between the Workflow Management Systems (WfMS) AiiDA, jobflow, and pyiron. It is based on three components: (1) a conda environment that specifies the software dependencies, (2) a Python module that contains the Python functions represented as nodes in the workflow graph, and (3) a workflow graph stored in the JavaScript Object Notation (JSON). The application of the PWD is demonstrated on three different workflows with different combinations of Python functions and external executables, which require interfacing using file-based communication, highlighting the universal applicability of the PWD. With the corresponding Python interface that we developed, users can export DAG-based workflows from one WfMS to the PWD and then import the PWD representation of the workflow with any of the supported WfMS.
After the import of the workflow, the user still has the option to adjust the input parameters of the workflow, adjust and add WfMS specific features, and assign computational resources to leverage HPC during the execution of the workflow. In the current version, the assignment of the computational environment is not included in the PWD as it is not expected that a user would use multiple WfMS on the same HPC cluster, but rather uses the PWD when transferring a workflow from one HPC cluster with a specific WfMS to a different HPC cluster with a different WfMS. In this case, the assignment of the compute environment changes based on the different HPC resources.

Future development directions of the PWD will focus on broadening its adoption and enhancing its capabilities:
\begin{itemize}
\item Engage a wider array of WfMS developers and scientific communities in the joint effort.
\item Enable connections to data handling frameworks like datatractor~\cite{datatractor}, 
and leverage the PWD to create containerized, portable versions of generalized workflows for both simulation and experiment.
\item Extend the PWD format to include standardized specifications for submitting workflows to standardized HPC resources, thereby simplifying execution across different infrastructures.
\item Transcend PWD's current limitation to DAGs by incorporating support for dynamic flow control elements like loops and conditional branching, enabling the representation of more complex scientific workflows.
\end{itemize}
Ultimately, the vision is to evolve the PWD towards a comprehensive schema capable of capturing all information necessary to define computational workflows, from initial setup to final results, beyond the field of materials science.
For this vision the key difference of the PWD in comparison to other workflow standardization efforts is the use of the Python programming language to define workflow nodes, which benefits from the wide adoption of the Python programming language in the scientific community and the direct transfer of data in memory, without requiring to store intermediate results in files. 

\section*{Author contributions}
\textbf{Jan Janssen}: Writing – original draft, Conceptualization, Investigation, Methodology, Software, Visualization, Project administration. \textbf{Janine George}: Writing – original draft, Methodology, Funding acquisition. \textbf{Julian Geiger}: Writing – original draft, Investigation, Software. \textbf{Marnik Bercx}: Writing – review \& editing, Methodology. \textbf{Xing Wang}: Writing – review \& editing, Investigation, Software. \textbf{Christina Ertural}: Writing – review \& editing. \textbf{Joerg Schaarschmidt}: Writing – review \& editing. \textbf{Alex Ganose}: Writing – review \& editing. \textbf{Giovanni Pizzi}:  Writing – review \& editing, Methodology, Funding acquisition. \textbf{Tilmann Hickel}: Writing – review \& editing, Funding acquisition. \textbf{Joerg Neugebauer}: Writing – review \& editing, Methodology, Funding acquisition.

\section*{Conflicts of interest}
The authors declare that they have no known competing financial interests or personal relationships that could have appeared to influence the work reported in this paper.

\section*{Data availability}
The Python implementation of the PWD \mintinline{python}{python_workflow_definition} including all the examples from the paper are available at Ref.~\cite{repository}.

\section*{Acknowledgements}
JJ, JS,  TH, and JN thank the German Federal Ministry of Education and Research (BMBF) for financial support of the project Innovation-Platform MaterialDigital (www.materialdigital.de) through project funding FKZ no: 13XP5094A,  13XP5094C, and 13XP5094E. Further JJ, TH and JN also acknowledge funding from the Deutsche Forschungsgemeinschaft (DFG) through the CRC1394 “Structural and Chemical Atomic Complexity – From Defect Phase Diagrams to Material Properties”, project ID 409476157 and the consortium NFDI-MatWerk under the National Research Data Infrastructure, NFDI 38/1, project ID 460247524. CE and JaG acknowledge the Gauss Centre for Supercomputing e.V. (https://www.gauss-centre.eu) for funding workflow-related developments by providing generous computing time on the GCS Supercomputer SuperMUC-NG at Leibniz Supercomputing Centre (www.lrz.de) (Project pn73da). JaG was supported by ERC Grant MultiBonds (grant agreement no: 101161771; Funded by the European Union. Views and opinions expressed are, however, those of the author(s) only and do not necessarily reflect those of the European Union or the European Research Council Executive Agency. Neither the European Union nor the granting authority can be held responsible for them.) JuG, MB, XW and GP acknowledge financial support from the NCCR MARVEL, a National Centre of Competence in Research, funded by the Swiss National Science Foundation (grant no
: 205602), and from the SwissTwins project, funded by the Swiss State Secretariat for Education, Research and Innovation (SERI). GP acknowledges financial support from the Open Research Data Program of the ETH Board (project ``PREMISE'': Open and Reproducible Materials Science Research).




\renewcommand\refname{References}

\bibliographystyle{rsc}
\bibliography{literature} 
\end{document}